# First-principles calculations of the phonon dispersion curves of H on Pt(111)

Sampyo Hong and Talat S. Rahman*

Department of Physics, Cardwell Hall, Kansas State University, Manhattan, KS 66506

Rolf Heid and Klaus Peter Bohnen

Forschungszentrum Karlsruhe, Institut für Festkörperphysik, D-76021 Karlsruhe, FRG

**Abstract**

We have calculated the surface phonon dispersion curves for H on Pt(111), using first-principles, total energy calculations based on a mixed-basis set and norm-conserving pseudopotentials. Linear response theory and the harmonic approximation are invoked. For one monolayer of H in the preferred adsorption site (fcc hollow) vibrational modes polarized parallel and perpendicular to the surface are found, respectively, at 73.5 meV and 142.6 meV, at the Γ point of the surface Brillouin zone. The degeneracy of the parallel mode is lifted at the zone boundaries, yielding energies of 69.6 meV and 86.3 meV at the M point and 79.4 meV and 80.8 meV at the K point. The dispersion curves for H adsorption at the hcp hollow site differ only slightly from the above. In either case, H adsorption has considerable impact on the substrate modes; in particular the surface mode in the gap in the bulk phonon spectrum (around M point) is pushed into the bulk band. For on-top H adsorption, modes polarized parallel and perpendicular to the surface have respective energies of 47.4 meV and 277.2 meV, at the Γ point. The former disperses to 49.1 meV and 59.5 meV at the M point and to 56 meV and 56.7 meV at the K point. The H vibrational mode polarized perpendicular to the surface shows little dispersion, in all three cases considered. Insights are obtained from the hybridization of the H and Pt electronic states.

# I. INTRODUCTION

The investigation of the vibrations of atoms and molecules at surfaces is important and interesting as they provide a direct link to the surface force fields which further disclose the nature of the bonding between surface atoms and how it differs from that between atoms in the bulk solid. Insights so gained about the details of the potential energy surface also contribute to the understanding of surface phenomena such as adsorption and diffusion which are critical ingredients of the industrially relevant process of catalysis. In general, atoms adsorbed on surfaces vibrate with frequencies and displacement patterns that depend on the symmetry of the adsorption site and the surface geometry. Surface vibrations have thus been used for the determination of adsorption site and adsorbate-substrate bonding strengths. While some of this information is already available from the measured and calculated vibrational frequencies of the modes at the Brillouin zone center ($\Gamma$ point), more detailed insights and information are offered by the dispersion of the phonon frequencies along the high symmetry directions in the two dimensional surface Brillouin zone. In particular, the characteristics of the dispersion curves speak to the nature of the adsorbate-substrate interactions and adsorbate-adsorbate lateral interactions, both direct and those introduced through the substrate. In this study we present the calculated phonon dispersion curves of a monolayer of H adsorbed on Pt(111) for several possible H adsorptions sites. Our goal is two-fold. First we are interested in finding how the adsorption of a light atom like H affects the dispersion of the substrate surface phonons. Such a study would tell us about the nature of the bonding between H and the surrounding Pt atoms, and between the Pt surface atoms. Second, distinct features in the phonon dispersion curves may help resolve lingering issues about H adsorption site on Pt(111) which has been the subject of much experimental and theoretical investigation for some time now.[1-25] In an earlier theoretical and experimental study of the surface phonon dispersion curves based on parameterized force constants[1] some conclusions have already been drawn about the effect of H adsorption on Pt surface force constants and on the frequency of the Pt Rayleigh mode. One conclusion is that while the force constants between the surface atoms on clean Pt(111) are different from the values in the bulk, adsorption of 1 ML of H restores these force constants to almost the bulk values. However, these studies based on empirical models were meant to provide an understanding of some controversial features observed in He atom-surface scattering data for the phonons of clean Pt(111) and did not address issues specific to the details of H adsorption on Pt(111). The focus was thus on modes with frequencies lower than that of the Rayleigh waves and the dispersion of modes with higher frequencies including the H vibrational modes were not presented.

After extensive debate, three decades of experimental and theoretical studies of H on Pt(111) have led to the conclusion that H adsorbs preferably in the threefold hollow site, forms an ordered (1x1) overlayer at saturation coverage, and hybridization occurs between the 1s states of H and the d bands of Pt. The lingering issue is that of the assignment of the vibrational modes observed for full coverage which also raises questions about the symmetry of the adsorption site. While the very early high resolution electron energy loss spectroscopy (HREELS) measurements of Baro et al.[3] had revealed only two modes, one at 68 meV and another at 153 meV, at the Brillouin zone center, subsequent HREELS measurements have displayed an additional mode at about 112 meV.[5,12] The presence of the third mode has rekindled the debate on the preferred site for H adsorption,[3-10] the polarization of the modes,[3,5,11,12] and whether H remains localized or delocalized on the surface.[12-16] The extent of anharmonicity of the H modes has also been discussed with the conclusion that while the mode polarized normal to the surface is reasonably harmonic, the one in the direction parallel to the surface is significantly anharmonic.[9,12,16] The issue of the adsorption site is all the more intriguing because in the thermal desorption data taken by Jacobi and coworkers[12] there is the onset of an additional peak for H coverage exceeding an estimated 0.75 ML. The diffusion barrier for H on Pt(111) has also been found to be small (~70 meV),[17] giving further testimony to the equivalence of the three adsorption sites: fcc-hollow, hcp-hollow, and on-top. Furthermore, the frequencies of H vibrations are found to be coverage dependent. Recent HREELS measurements find modes at lower coverage ($\theta < 0.75$) at 31 meV and 68 meV, which they designate as polarized normal and parallel to the surface, respectively.[12] For higher coverage ($0.75<\theta<1$) the modes with polarization normal to the surface appear to be at 112 meV and 153 meV,[5,12] while that polarized in the surface plane is at 68 meV. Related theoretical calculations[12] assign the modes at 68 meV and 153 meV to vibrations localized around an equilibrium position in the three fold hollow site. The mode at 112 meV is not easily accounted for in these calculations, for obvious reasons. The symmetry of the surface requires that the two in-plane modes be degenerate at the Brillouin zone center, if H occupies a single adsorption site. The third modes has thus been explained as either an overtone, a combination loss,[5] or as a mixed mode arising from H delocalized over the fcc and hcp hollow sites.[12] On the other hand, a more complex adsorption scheme could lead to some interesting conclusions about H vibrational frequencies. To explain details in the analysis of photoemission data, a two-state adsorption model was proposed,[2,4] but subsequently argued to be incompatible with photoemission data.[8] The proposal of subsurface H has reemerged in recent times,[16,18] although clear experimental evidence is yet to be provided.

In this study we consider first the case of 1 ML coverage of H on Pt(111) and compare the structural and dynamical properties of the system for the three surface adsorption sites: fcc-hollow, hcp-hollow, and on-top. Our calculations are based on density functional perturbation theory (DFPT) based on the linear response theory and the harmonic approximation.[26] For a possible explanation of the observed frequency of the third mode at the Brillouin zone center, we consider also the case of subsurface H adsorption for 2 ML coverage. This higher coverage is not unreasonable since the exact knowledge of coverage from experiments performed for saturation coverage is not easy to obtain, particularly for an adsorbate like H.

Below we provide some details of our calculations in Section II, followed by a description of our model system in section III. The results for 1 ML coverage are presented in Section IV with discussion of the calculated structural properties followed by that of the phonon dispersion curves. The case of subsurface adsorption for 2 ML coverage is presented in Section V together with summary of our conclusions.

## II. THEORETICAL DETAILS

The total energy of the fully relaxed systems was calculated using the density functional theory within the pseudopotential method in a mixed basis representation.[27] To represent ion-electron interaction, a norm-conserving pseudopotential with scalar-relativistc correction was used in the local density approximation (LDA). The Pt pseudopotential includes, except spin-orbit coupling, all relativistic corrections up to order $\alpha^2$, where $\alpha$ is the fine structure constant. The electron-electron interaction[28] was represented by a Hedin-Lundqvist form of the exchange-correlation functional.[29] For the valence states of Pt and H local functions of d and s type with radial cutoffs of 2.1 a.u. and 0.7 a.u., respectively, were applied. Kinetic energy cutoff for plane waves was 16.5 Ryd. Integration over an irreducible Brillouin-zone were carried out using 42 special kpoints. A Fermi level smearing of 0.014 Ryd was employed.[30] Structural relaxations were carried out until the forces, calculated using the Hellmann-Feynman theorem, on all atoms were less than 0.001 Ryd/a.u.

The calculations of the vibrational frequencies and their eigenvector were performed using linear response theory within a perturbative approach[31] in a mixed-basis representation of the wave functions. Phonon dispersion curves were obtained by standard Fourier interpolation method using a (6x6) q-point mesh. Surface force constants were merged with bulk force constants and an asymmetric bulk slab of 100 layers was added to obtain projected bulk phonon modes[32] using standard lattice dynamical methods. An (8x8x8) q-point mesh was used for the calculation of the bulk phonons. Vibrational modes were labeled as surface modes if their eigenvectors contained contributions larger than 20% of the displacements of the atoms in the top two Pt layers.

## III. MODEL SYSTEMS

In Fig. 1(a) the top view of a close-packed arrangement of spheres representing a segment of the fcc(111) surface is shown. The large white circles are Pt atoms. The three adsorption sites, fcc, hcp and on-top are marked, respectively, as A, B, and C in the figure. 1 ML coverage of H in the fcc site, for example, would mean occupation of all sites labeled A in Fig. 1(a). Beneath the points B there are Pt atoms in the second layer, while underneath A the substrate atoms are in the third layer, consistent with fcc stacking of solids. Since these three adsorption sites have been the subject of much discussion and since the calculated adsorption energy for H in each of these sites is close to that measured experimentally,[10] we have carried out calculations of the dispersion of the surface phonons for each of them. The two dimensional Brillouin zone for fcc(111) surface is shown in Fig. 1(b).

For the calculations a total of 9 layers of Pt atoms and one H atom on each side were used for a slab with inversion symmetry. The surface (1x1) unit-cell used to simulate 1ML H adsorption is shown in Fig. 1(a) as a parallelogram.

# IV. RESULTS AND DISCUSSIONS

## A. Structural Relaxations

The calculated relaxed geometrical structures for 1 ML of H adsorbed in each of the three sites on Pt(111) are summarized in Table I. In agreement with previous theoretical calculations,[9] we find the height of the H atoms above the Pt surface for fcc, hcp, and on-top site adsorption to be 0.93 Å, 0.93 Å, and 1.56 Å, respectively, with H-Pt bond lengths of 1.85 Å, 1.85 Å, and 1.56 Å. We find distance between H in the hcp-site and the second layer Pt atoms directly below to be 3.3 Å. Our calculated adsorption energies for 1 ML H coverage are 0.41 eV (fcc), 0.37 eV (hcp), and 0.31 eV (on-top) showing a slight preference for fcc site adsorption, and in close agreement with experimental results (0.23 ~ 0.47 eV) for a range of coverage.[2, 19 - 23] In a detailed theoretical study[9] in which density functional theory calculations were carried out for the H/Pt(111) system with both LDA and the generalized gradient approximation (GGA), as well as, scalar relativistic and spin-orbit corrections, the preferred adsorption site for H on Pt(111) was also predicted to be the fcc hollow. Interestingly, Olsen et al[9] find GGA to provide better agreement with experimental values of the adsorption energies while LDA is found to overestimate them: 0.75 eV ~ 0.81 eV (fcc) and 0.71 eV ~ 0.8 eV (top) depending on whether full relativistic or scalar relativistic corrections were used. In a more recent study, Legare also using GGA finds adsorption energies in the range of 0.43 eV – 0.48 eV for the two hollow site adsorption.[18] The differences in the adsorption energies reported here from those in Ref. 9 may have to do with the extent of self-consistency, number of layers in the super cell, and other details employed in the two sets of calculations. To remove ambiguities in the calculated values of the adsorption energy, we note that our values are with respect to those for H in the molecular state in gas phase. We also note that in all calculations the adsorption energies for the three sites are found to be close to one another, pointing to the flatness of the potential energy surface for H on Pt(111).

The changes in Pt interlayer separations are also noted in Table I. Here $\Delta_{ij}$ is the percentage change in the spacing between the layers labeled i and j. The Table shows that H causes an outward relaxation of the Pt surface layer of 2.3% (on-top), 2.7% (fcc) to 3.7% (hcp).[24] However, noticeable relaxations are confined only to the topmost layer, reflecting the compactness of the (111) surface. The calculated surface relaxations are in agreement with experimental observations. Note also that clean Pt(111) has been found to relax only marginally outwards.[25] In general, the results in Table I indicate that the relaxation of the Pt surface layer induced by H adsorption is similar for the three adsorption sites.

## B. Surface phonon dispersion curves

The surface phonon dispersion curves for 1ML of H on Pt(111) have been calculated along two high symmetry directions, $\Gamma – M$ and $\Gamma – K$, in the surface Brillouin zone shown in Fig. 1(b). In the dispersion curves displayed in Fig. 2, grey solid lines correspond to bulk-projected modes, the dark filled-circles are the Pt substrate surface modes (lower panel before the break in the vertical axis) or the vibrations of H atoms (upper panel after the break), and the white circles represent the data from He-atom surface (HAS) measurements.[1] Frequencies of prominent Pt and H surface phonons at the high symmetry points in the Brillouin zone are tabulated in Tables II and III, respectively. It should be noted that an asymmetric slab filing was performed which gave two types of surfaces in the model system: one H-covered and the other Pt bulk-terminated. Each of these surfaces contributed one set of curves well separated from the bulk-projected band. The set of dispersion curves in Fig. 2 thus show a grey line below the bulk-projected band, originating from the bulk-terminated Pt surface which should be ignored as it is an artifact of our calculational set up and not a part of the physical system under consideration. Note that for discussions purposes we number the surface modes from the bottom to the top of the band. The first (lowest energy) surface mode in Figs. 2(a) – 2(c), at the zone boundaries, are thus the Rayleigh Waves (RW) which are easily identifiable since they are generally well separated from the bulk-projected bands and usually have a vertical polarization at the Brillouin zone boundaries.

At the M point, the dispersion curves for H adsorption in the fcc-site [Fig. 2(a)], display three Pt substrate surface modes with frequencies 9.6 meV, 21.9 meV, and 22.3 meV (Table II). The RW (9.6 meV) consists of purely vertical vibrations of the atoms in the top Pt layer. The second (21.9 meV) and the third (22.3 meV) modes comprise purely longitudinal displacements of the atoms in the first and second Pt layers. At the K point, we find four surface modes with frequencies 9.7 meV, 13.3 meV, 17.4 meV, and 19.6 meV. The frequency of the RW (9.7 meV), with pure vertical polarization, is in good agreement with that in the HAS measurements.[1] The second substrate surface mode at K (13.3 meV) is mainly of vertical character with a weak parallel component. The three topmost Pt layers take part in this mode such that there is a vertical displacement of Pt atoms in the second layer while the atoms in the first and third Pt layers vibrate in the direction parallel to the surface. The third substrate mode (17.4 meV) is polarized purely in the surface plane, however, its amplitude extends below to five Pt layers. The fourth mode (19.6 meV) is a surface resonance and its amplitude penetrates deep into the bulk. The dispersions of the H vibrational modes are represented by the top three curves in Fig. 2(a). The two lower curves arise from in-plane vibrations, while the top curve is the result of vertical displacements of the H atom. The parallel vibrational modes show substantial dispersion with a maximum split of 16.6 meV at the M point. The upper branch consists of longitudinal modes, while the lower one is the shear-horizontal mode, along both the Γ- M and Γ- K directions. At the K point, the two H parallel modes are almost degenerate with a maximum difference in frequency of only 1.4 meV.

In Fig. 2(b) the corresponding dispersion curves for H adsorption in the hcp site are shown. They appear to be very similar to those discussed above for fcc-site adsorption, except that some substrate surface modes are slightly softened. Also, the dispersion of the H modes display a larger split in the frequencies of the parallel mode at the M point (20.4 meV), as compared to those for fcc-site adsorption [Fig. 2(a)]. A larger difference from the dispersion curves above for the two hollow-site adsorptions is exhibited in Fig. 2(c) by those calculated for on-top adsorption. For H vibrations parallel to the surface, at the Γ point, the frequency (47.4 meV) is much smaller than those obtained for the hollow-sites, while the frequency of the one polarized in the direction normal to the surface (277.2 meV) is much larger. The splitting of the frequencies of the modes polarized parallel to the surface at the M point (10.4 meV) is also much smaller than that for the hollow-site adsorption.

In addition to the differences in the dispersion of the H modes, the calculated dispersion of the Pt surface modes shows a striking difference for the hollow-site and top-site adsorptions. The third substrate surface mode continues to lie in the "stomach" gap, for H top-site adsorption, as for clean Pt(111), while it gets pushed into the bulk band for hollow-site adsorption. Actually, it is interesting to quantify the impact of H adsorption on Pt surface phonon dispersions, since it is so often assumed that the substrate phonons remain unchanged in the presence of an adsorbate as light as H.[1] To get a better estimate of the shifts in the frequencies on H adsorption, the calculated surface phonon dispersion curves for clean Pt(111) are displayed together with experimental values (white circles) in Fig. 2(d). While detailed analysis of the results in Fig. 2(d) can be found in a recent review article,[33] we comment here on the longitudinal resonances which were found in He scattering experiments but not reproduced in theoretical calculations. These are the sets of experimental points in Fig. 2(d) above the Rayleigh mode along both the Γ- M and Γ- K directions. Our calculations do not reveal such modes either for clean Pt(111) or for that with a monolayer of H adsorbed on it, as is clear from the four sets of dispersion curves in Fig. 2. Note that in Ref. 1 also the longitudinal modes were found to disappear once H was adsorbed on Pt(111).

In comparing the dispersion curves in Fig. 2, we find that the surface modes for clean Pt(111) undergo a few changes on H adsorption: two RW or RW-like modes on the clean surface, near the zone boundaries and along the M – K direction, are replaced by only one in the dispersion curves for the H-covered surface. This is because the first mode which is vertically polarized (RW), at the zone boundaries, is softened and the next two surface modes (polarized in the surface plane) are stiffened such that they are pushed above into the bulk hand, on H adsorption in the hollow-site (Table II). This trend is repeated for the top-site adsorption, albeit somewhat weakly. Clearly H adsorption alters the force field of the surface atoms and consequently the surface phonon dispersion curves. Below we discuss some of the high lights of the Pt(111) surface force constant changes brought about on H adsorption.

### C. Force constant changes

The calculated force constant matrix for the Pt(111) surface atoms with 1 ML coverage of H exhibit substantial changes from the values on the clean Pt(111) surface. These changes are found to be extensive and a full representation of the tables of the force constants between all surface atoms for the three adsorption sites would be quite cumbersome and of limited interest. To fully appreciate the changes in the strength of coupling between any two atoms we also need to take into account all components in the force-constant matrix $\Phi_{ij}$ between atoms in layers i and j. For this purpose, the convenient concept of an average coupling-strength <I> for a bond length σ is introduced.[28] It is defined as $<I> \equiv \sqrt{\frac{1}{3}\sum_{ij}\Phi_{ij}^2(\sigma)}$ .

In Table IV values of <I> and its percentage changes from the bulk values (100 ∗ (<I>$_\sigma$ - <I>$_{bulk}$)/<I>$_{bulk}$) are presented. Consistent with previous results,[1] we find a remarkable weakening (33%) of the lateral interaction between Pt atoms in the top layer on the clean surface from the bulk value. The Table shows that on-top H adsorption produces only a small change in the Pt surface in-plane force constants from the values on the clean surface, while for hollow site adsorptions these force constants undergo a stiffening of about 26%. The stiffening of these force constants leads to shift towards higher frequencies of the surface mode in the stomach gap, in the dispersion curves in Figs. 2(a) and 2(b) (fcc: 21.9 meV, hcp: 21.3 meV). This is understandable since this stomach gap mode consists mainly of in-plane motion of the Pt surface atoms. Since the frequency of this mode remains unshifted for top-site adsorption, this feature can serve as a signature of the H adsorption site on Pt(111). Table IV also shows that the interlayer average-coupling strength hardly changes for the clean Pt(111) surface with respect to bulk, but it softens by 28% for hcp and on-top H adsorption and by 18% for fcc adsorption. The changes in the average force constants are also in qualitative agreement with the trends in the calculated structural relaxations.

### D. Local density of states and charge density distribution

An important conclusion from the above is that H adsorption in the hollow sites reverses the change in the lateral force constants between the Pt atoms in the top layer from large softening on the clean surface to stiffening on the adsorbate covered surfaces. To obtain a deeper understand of the physics behind these observations, we turn to a full study of the surface electronic structure of H-covered Pt(111). For this we calculate the local density of states through projections of the total wavefunction onto atoms of interest within the Wigner-Seitz spheres around them. The Wigner-Seitz radii for the H and Pt atoms are taken to be 0.5Å and 1.35Å, respectively. In Figs. 3 and 4 we compare the resulting local density of states (LDOS) of the Pt atoms in the bulk [Figs. 3(a) and 4(a)] with those for the atoms in the topmost layer of clean Pt(111) [Figs. 3(b) and 4(b)], and those with H adsorbed in the top site [Figs. 3(c) and 4(c)], the hcp site [Figs. 3(d) and 4(d)] and the fcc site [Figs. 3(e) and 4(e)]. The LDOS for the s-band of H atoms are also shown in appropriate figures. As expected, the LDOS for the Pt atoms on the clean surface display features which are different from those in the bulk. In particular the d-band is found to contract for the surface atoms, and the LDOS is reduced in the lower energy region. More precisely, from Figs. 3(a), 3(b), 4(a), and 4(b) we find that the three Pt d-bands, $d_{xz}$, $d_{yz}$, and $d_{z^2}$ are most affected by the creation of the surface and lose states in the energy range -7 eV to -4 eV. However, the peaks in the LDOS around $E_F$ of the Pt atoms on the clean surface appear to be similar to those of the atoms in the bulk. For top-site H adsorption, shown in Figs. 3(c) and 4(c), the lost bonds of the Pt atoms on the clean (111) surface are recovered partly through the hybridization of the 1s states of H with the $d_{z^2}$ states of Pt. There is a remarkable depletion of the $d_{z^2}$ states near $E_F$ and several new peaks appear both below (5 – 7 eV) and above (~2 eV) the Fermi level. Figs. 3(c) and 4(c) also reveal that the $d_{xz}$, $d_{yz}$ and $d_{x^2-y^2}$ states of Pt are not significantly affected by H adsorption in the top site.

The effect of H adsorption in the hollow sites on the LDOS of Pt surface atoms is significantly different from that for top-site adsorption. Figs. 3(d), 3(e), 4(d), and 4(e) attest to the contrast with the LDOS in Figs. 3(c) and 4(c), particularly in the hybridization of the H and Pt states. Additional peaks with contributions from $d_{xz}$ and $d_{x^2-y^2}$ states of Pt appear between 8 eV and 6 eV below $E_F$, in Figs. 3(d) and 3(e) for hollow site adsorption. These figures also show the strong hybridization of the Pt states with those of H. Clearly the Pt 6s states also participate in the hybridization when H adsorbs in the hollow sites.

The charge density distribution corresponding to new peaks in the region -8 eV to -6 eV in Figs. 3(e) and 4(e) is shown in Fig. 5. The plane chosen for the side-view display in the figure passes through the points labeled C-A in Fig. 1(a). The upper center of the "chicken-leg" shape is located at the center of the adsorbed H (in fcc hollow), and the other is in the outer shell of the Pt surface atom close to H. The asymmetry in the charge distribution points to the difference in the contributions coming from H and Pt atoms. However, it is clear that the bonding charges are distributed along the line that connects H to the Pt atoms, and points to the strong H-Pt covalent bonding.

### E. Subsurface adsorption

As discussed in Introduction, the possibility of subsurface adsorption has been the subject of several recent publications. Furthermore, it appears that in experiments the saturation coverage may correspond to more than 1 ML which begs the question of where the other monolayer goes. We have carried out calculations for several configurations for H coverage of 2 ML in which 1 ML remains on the surface (fcc-site) and 1 ML is absorbed into the surface, between the first and second Pt layer. Such a configuration may be feasible as H is known to go subsurface on several metal surfaces (for example, Pd).[16] In the case that the subsurface monolayer is absorbed in the octahedral position below the top layer as shown in Fig. 6, our calculations for the Brillouin zone center produces two additional frequencies 50 meV and 106 meV. The wanted mode at 106 meV is the vertical vibration of the subsurface H with a small contribution from the vertical motion of the on-surface H. The mode at 50 meV consists of displacements of the sub-surface H in the direction parallel to the surface. The in-plane and vertically polarized modes arising from the on-surface H atoms have frequencies of 72 meV and 148 meV, respectively.

## V. CONCLUSION

We have calculated the phonon dispersion curves for H on Pt(111), using first-principles, total energy calculations based on a mixed-basis set and norm-conserving pseudopotentials. Linear response theory and the harmonic approximation are invoked. At 1 ML coverage of H on Pt(111), we find the preferred adsorption site to be the fcc hollow, at a height of 0.9 Å, with a binding energy of 0.41 eV, although we cannot rule out adsorption at the hcp-site. In fact, our results show very slight difference in the characteristics of the phonon dispersion curves for H adsorption in the two hollow-site.

In both cases, we find that H adsorption alters the force field of the surface through hybridization between the 1s states of H and $d_{xz}$, $d_{x^2-y^2}$, and 6s states of the Pt surface atoms in such way that the lateral interaction between the Pt atoms on the H-covered surface undergoes stiffening rather than softening which is exhibited by the clean Pt(111) surface. As a result, vertically polarized surface modes are lowered in energy while those polarized parallel to the surface are raised in energy. This trend is present in the top-site H adsorption in which hybridization occurs between H 1s and Pt $d_{z^2}$ states, however, it is weaker than that for hollow-sites.

For H vibrations, we find the in-plane polarized mode at 74 meV, and the vertically polarized one at 143 meV, at the Γ point, for the fcc-site adsorption. The corresponding frequencies for the hcp-site adsorption are 67 meV and 144 meV, while for the on-top site adsorption, they are 47 meV and 277 meV. The H mode polarized perpendicular to the surface shows little dispersion along the two high symmetry directions chosen in this study. The modes polarized parallel to the surface show some dispersion, particularly at the M point, and for hollow site adsorption. The frequency of the Pt Rayleigh wave, at the zone boundaries, is reduced on H adsorption in all three cases. The Pt surface mode in the stomach gap disappears into the bulk band for hollow site H adsorption and remains unshifted if H adsorbs on-top. These dispersion curves, however, do not reveal a mode around 112 meV as has been reported in experiments. Our calculation at the Γ point for the subsurface configuration in which 1 ML remains on the surface (fcc-site) and the subsurface monolayer is absorbed in the octahedral position below the top layer, produces the mysterious mode at 106 meV which is the vertical vibration of the subsurface H with a small contribution from the vertical motion of the on-surface H. The two-state adsorption model based on the subsurface location appears to be possibility.[18] We await further experiments particularly on the dispersion of the surface phonons of the H-Pt(111) system to provide further answers to the issue of H on Pt(111).

## ACKNOWLEDGEMENTS

The work was supported in part by the US National Science Foundation, Grant CHE-0205064. Computations were performed on the multiprocessors at NCSA, Urbana, and at Forschungszentrum, Karlsruhe. TSR also acknowledges the support of the Alexander von Humboldt Foundation and thanks her colleagues at the Fritz Haber Institut, Berlin and at the Forschungszentrum, Karlsruhe for their warm hospitality. We are grateful to Stefan Badescu and Karl Jacobi for getting us interested in the subject and for helpful discussions.

---

[*]Corresponding author. E-mail address: rahman@phys.ksu.edu; WWW homepage: http://www.phys.ksu.edu/~rahman/; FAX: 785 532 6806.

# Figures

FIG. 1. (a) The three probable adsorption sites for H on Pt(111): fcc hollow (A), hcp hollow (B), and on-top (C); (b) the surface Brillouin zone for the fcc(111) crystal.

FIG. 2. Surface phonon dispersion curves for one monolayer H on Pt(111): (a) fcc  (b) hcp  (c) on-top adsorption sites compared with that for clean Pt(111) in (d).

FIG. 3.  Calculated local density of states: $d_z^2$, $d_{x^2-y^2}$, and 6s states of Pt atoms in (a) the bulk, (b) the top layer of clean Pt(111), (c) – (e) the top layers of Pt(111) with 1s state of H for the three possible H adsorption sites.

FIG. 4.  Calculated local density of states: $d_{xz}$, $d_{yz}$, and $d_{xy}$ of Pt atoms in (a) the bulk, (b) the top layer of clean Pt(111), (c) – (e) the top layers of Pt(111) with 1s state of H for the three possible H adsorption sites.

FIG. 5. Charge density distribution for H adsorbed at the fcc hollow site.

FIG. 6. The configuration considered for subsurface adsorption of H in the octahedral sites below the top layer: (a) top view; (b) side view. The small white circles are H atoms, and the large white, gray, and black circles represent Pt atoms in the first, second, and third layer, respectively. The octahedral adsorption site is directly below the fcc site.

# Tables

TABLE I. Comparison of calculated surface geometry and interlayer relaxations for one monolayer of H on Pt(111) with experiments.

| Surface geometry | Theory | | | Experiment |
|---|---|---|---|---|
| Adsorption site | fcc | hcp | on-top | Three-fold hollow site |
| Adsorbate height | 0.93 Å | 0.93 Å | 1.56 Å | 0.58 ~ 1.15 Å [a] |
| $\Delta_{12}$ | +2.7 % | +3.7 % | +2.3 % | 0.7 ~ 2.7 % [b] |
| $\Delta_{23}$ | +0.2 % | +0.1 % | +0.4 % | |
| $\Delta_{34}$ | -0.5 % | -0.5 % | -0.3 % | |
| Binding energy | 0.41 eV | 0.37 eV | 0.31 eV | 0.23 ~ 0.47 eV [a] |

[a] Reference 2, 19 - 23.

[b] Reference 24.

TABLE II. Calculated frequencies (in meV) and polarization (P: parallel, V: vertical, M: mixed) of Pt surface phonon modes.

| wave-vector | fcc | hcp | on-top | experiment | clean |
|---|---|---|---|---|---|
| M | 9.6 (V) | 9.3 (V) | 9.8 (V) | | 11.0 (V) |
|   |          | 11.9 (P) | 11.3 (P) | | 11.5 (P) |
|   | 21.9 (P) | 21.3 (P) | 20.1 (P) | | 19.3 (M) |
|   | 22.3 (P) | 22.1 (P) |          | | 22.4 (P) |
| K | 9.7 (V) | 9.0 (V) | 9.8 (V) | 9.3 (V) | 11.1 (M) |
|   | 13.3 (M) | 13.2 (M) | 12.7 (M) | | 11.8 (V) |
|   |          | 17.1 (P) | 16.3 (P) | | 14.3 (P) |
|   | 17.4 (P) | 17.5 (P) | 16.6 (P) | | 16.1 (M) |
|   | 19.6 (P) |          |          | | 19.6 (P) |
| 0.65 Γ-M | 7.7 | 7.5 | 7.7 | 7.7 | |
| 0.81 Γ-K | 8.5 | 8.3 | 8.5 | 8.3 | |

TABLE III. Calculated frequencies (in meV) and polarization (P: parallel, V: vertical, M: mixed) of H vibrational modes.

| wave-vector | polarization | fcc | hcp | on-top |
|---|---|---|---|---|
| Γ | P | 73.5 | 67.3 | 47.4 |
|   | V | 142.6 | 143.9 | 277.2 |
| M | P | 69.6 | 64.2 | 49.1 |
|   |   | 86.3 | 84.6 | 59.5 |
|   | V | 138.4 | 139.6 | 276.2 |
| K | P | 79.4 | 75.0 | 56.0 |
|   |   | 80.8 | 77.2 | 56.7 |
|   | V | 137.3 | 137.6 | 276.1 |

TABLE IV. Average force constant matrix $\Phi_{ij}$ for Pt atoms in the topmost layer.

| | in-plane (1NN) $\Phi_{ij}$ ( i: 1$^{st}$ Pt layer  j: 1$^{st}$ Pt layer) (%: change from Bulk) | interlayer $\Phi_{ij}$ ( i: 2$^{nd}$ Pt layer  j: 1$^{st}$ Pt layer) (%:change from Bulk) | interlayer $\Phi_{ij}$ ( i: H  j: 1$^{st}$ Pt layer ) |
|---|---|---|---|
| fcc | 42658.6 (+27%) | 27788.1 (-18%) | 27063.0 |
| hcp | 42449.2 (+26%) | 24443.8 (-28%) | 27604.2 |
| on-top | 25826.1 (-23%) | 24647.3 (-28%) | 163806.5 |
| clean | 22675.6 (-32%) | 33083.4 (-3%) | |

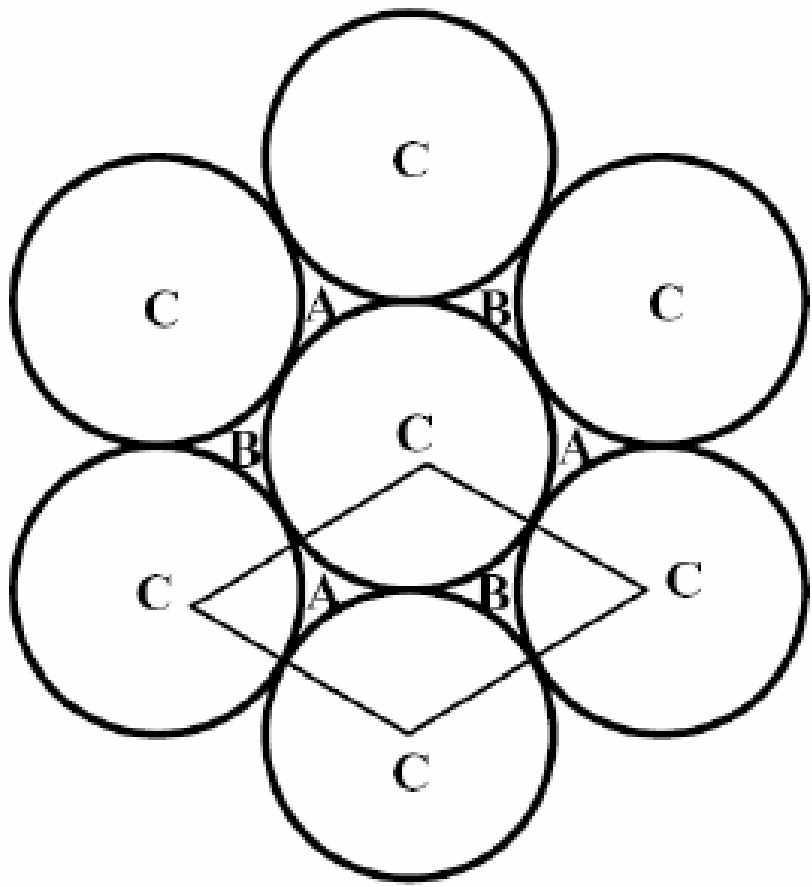

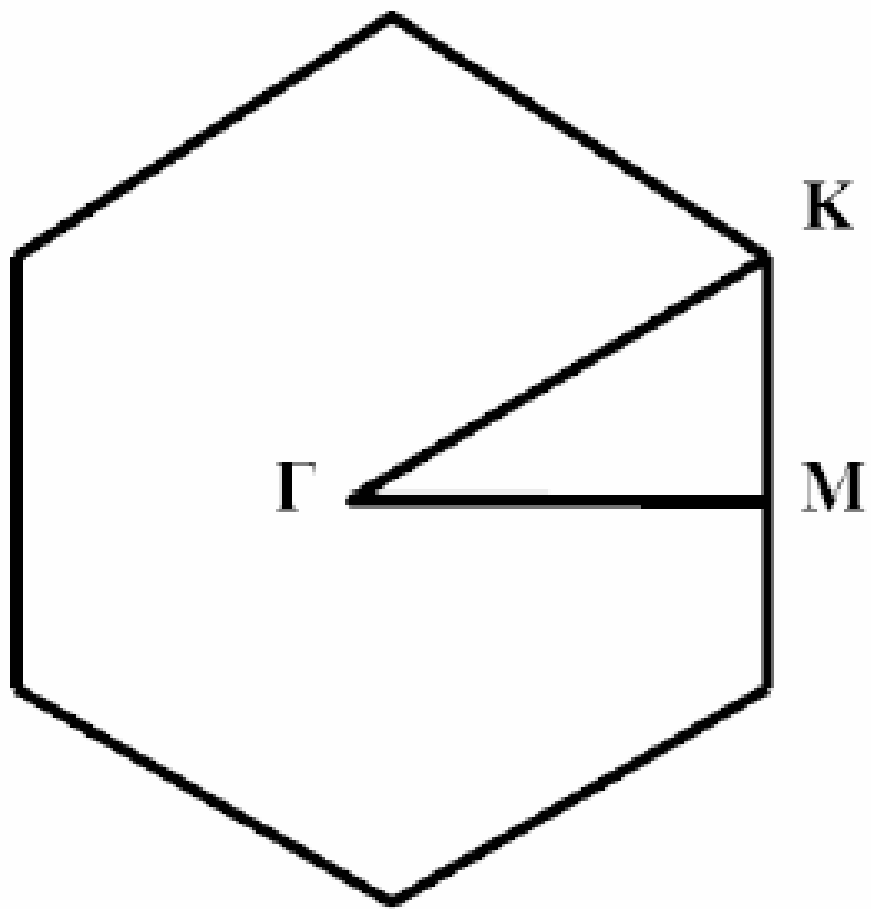

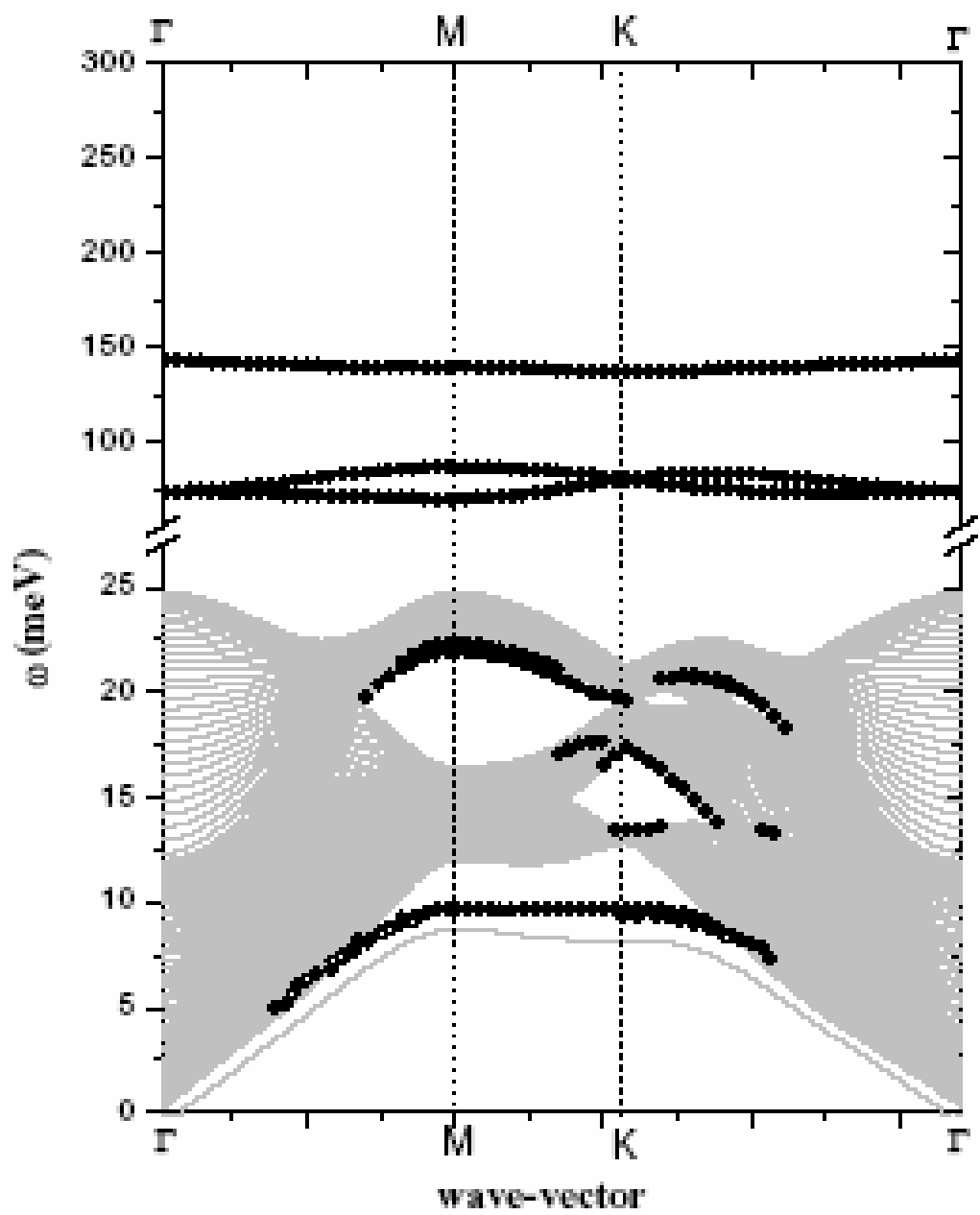

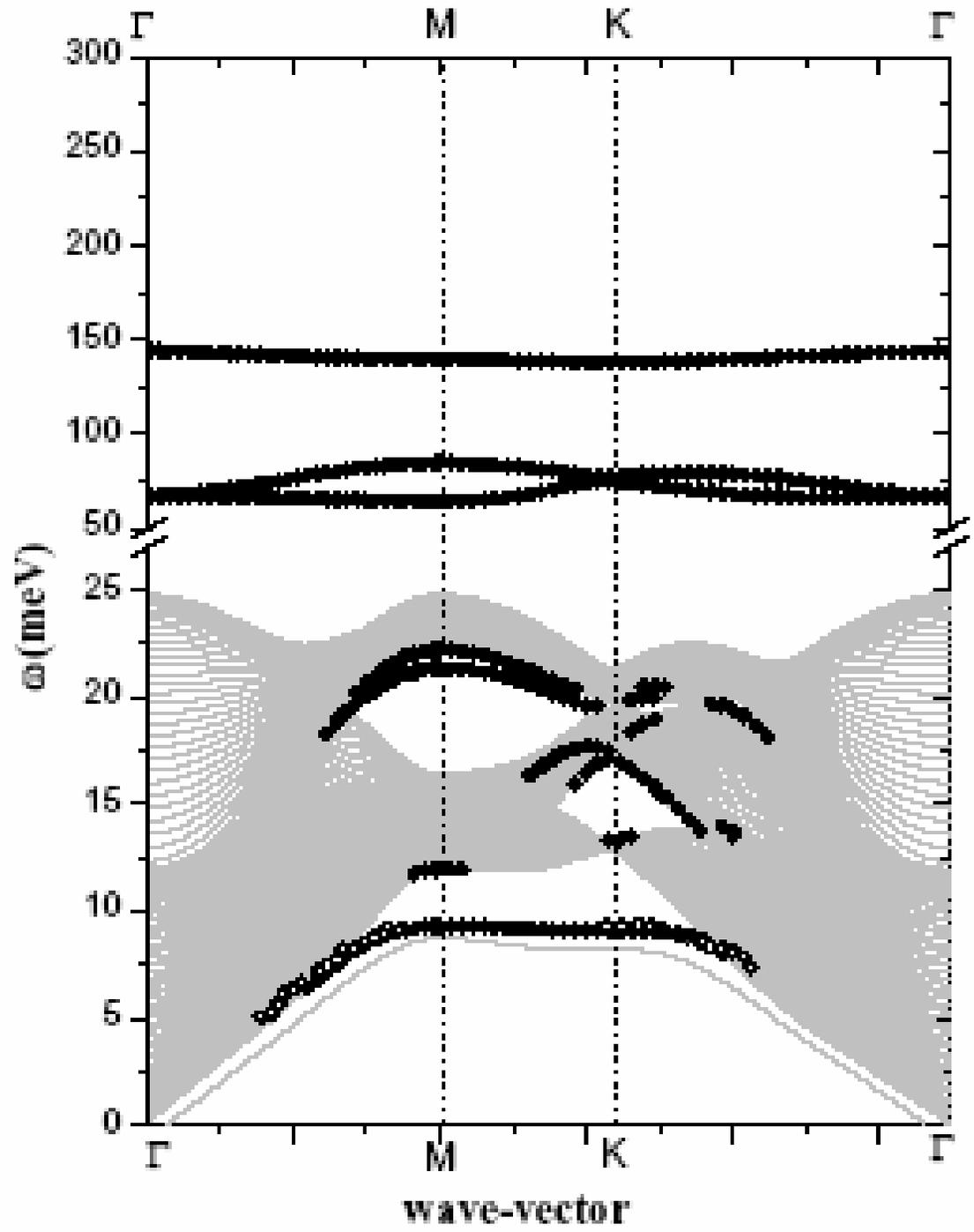

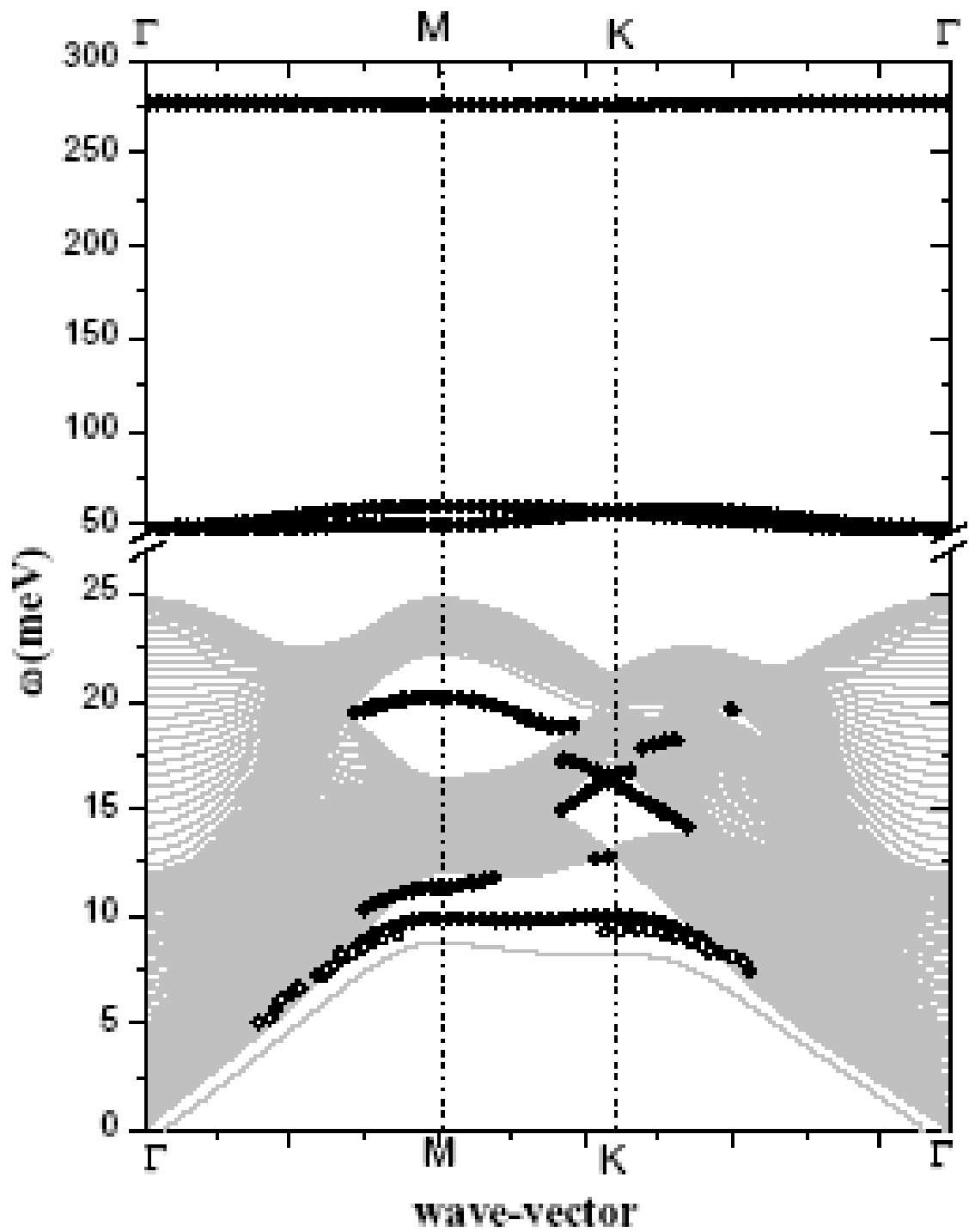

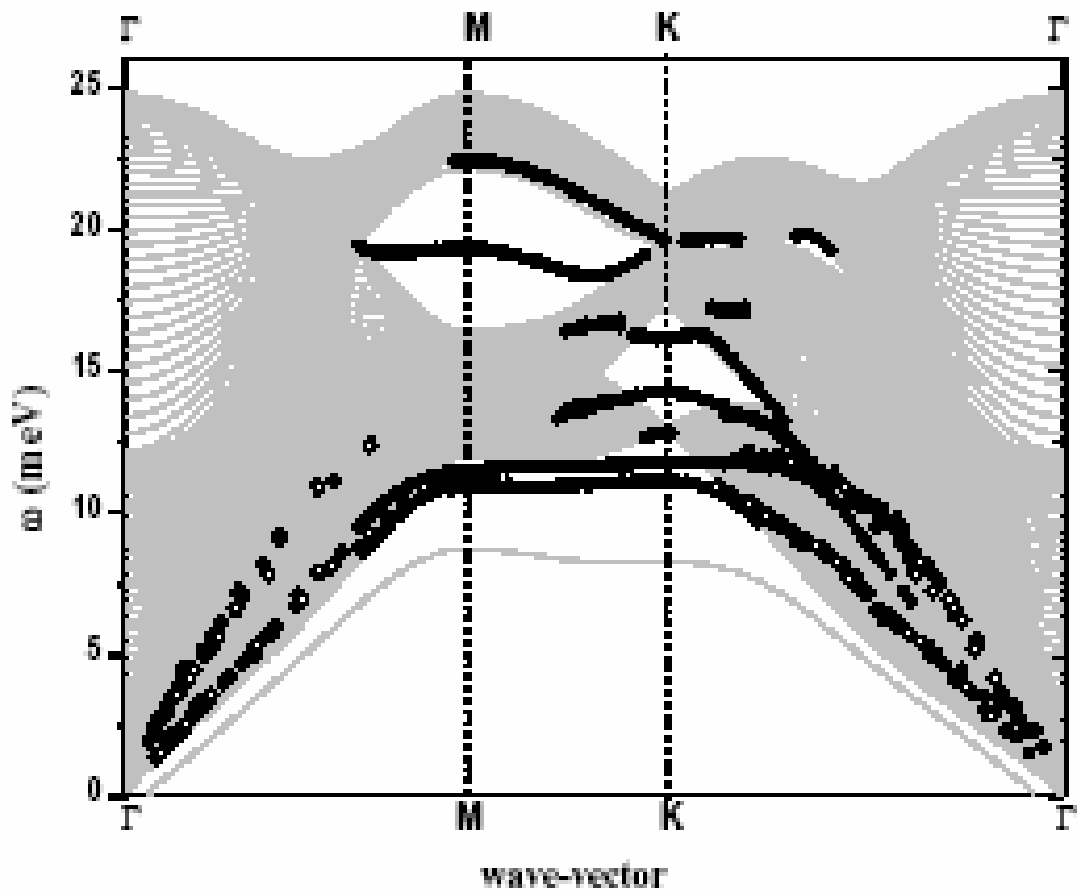

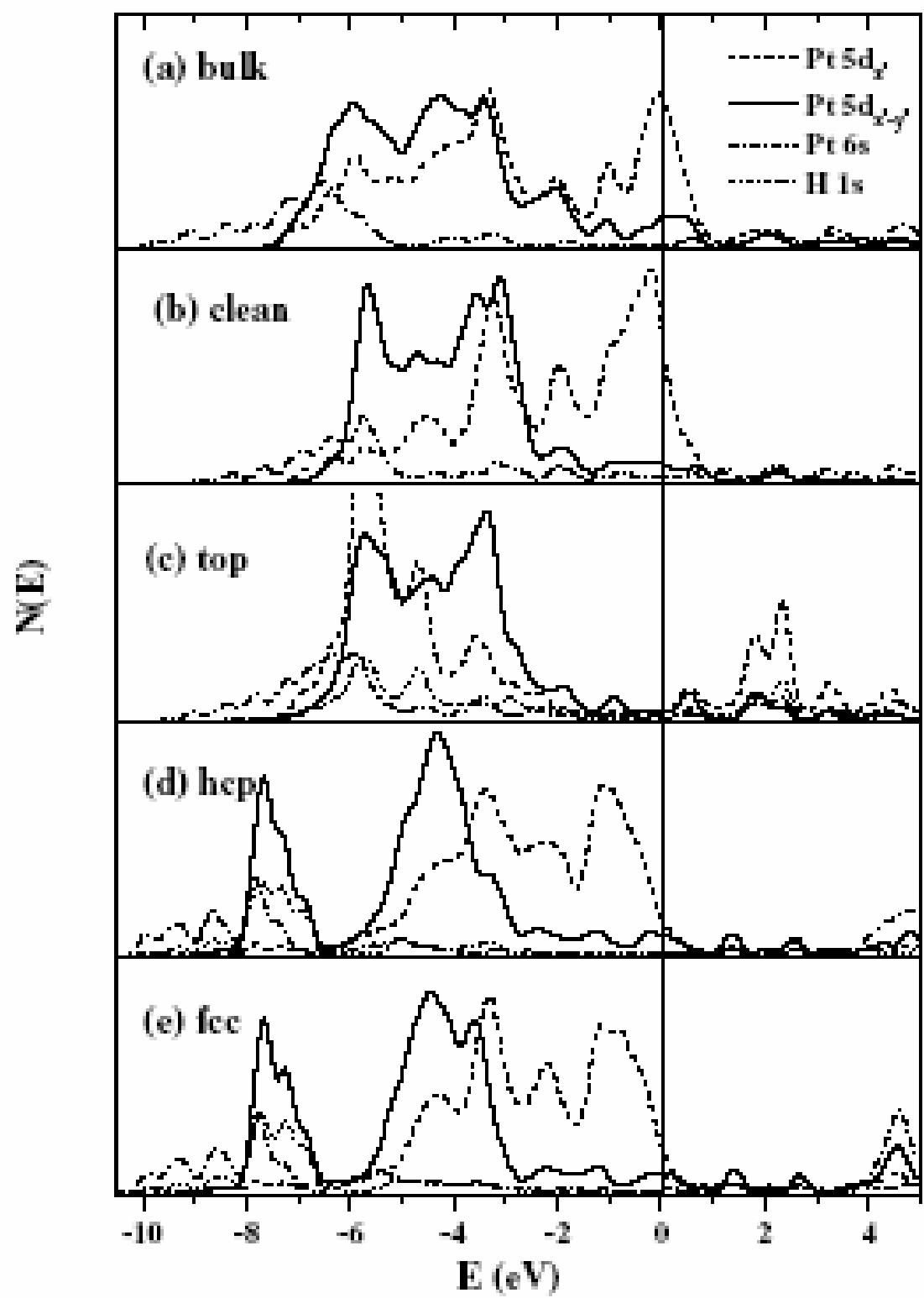

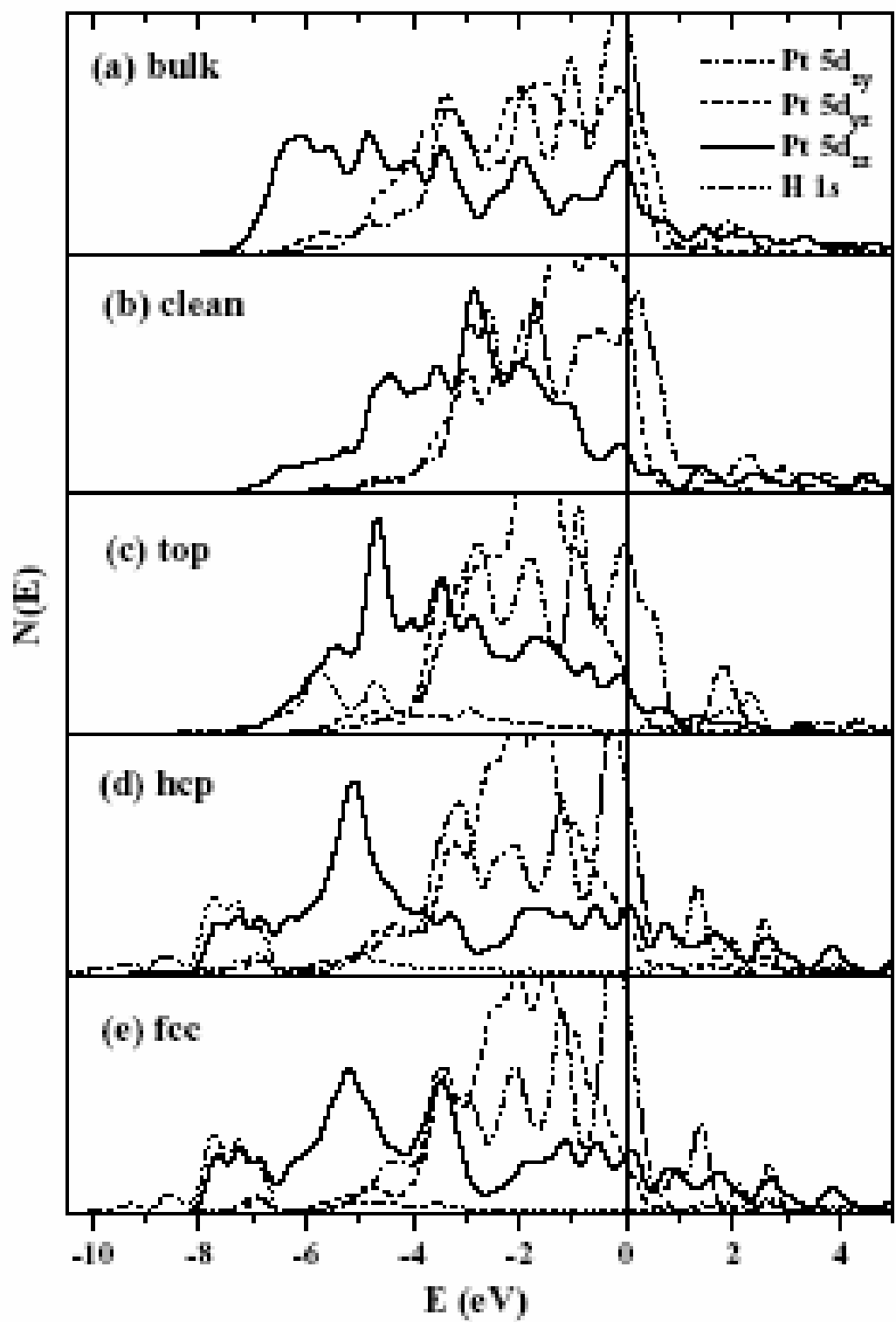

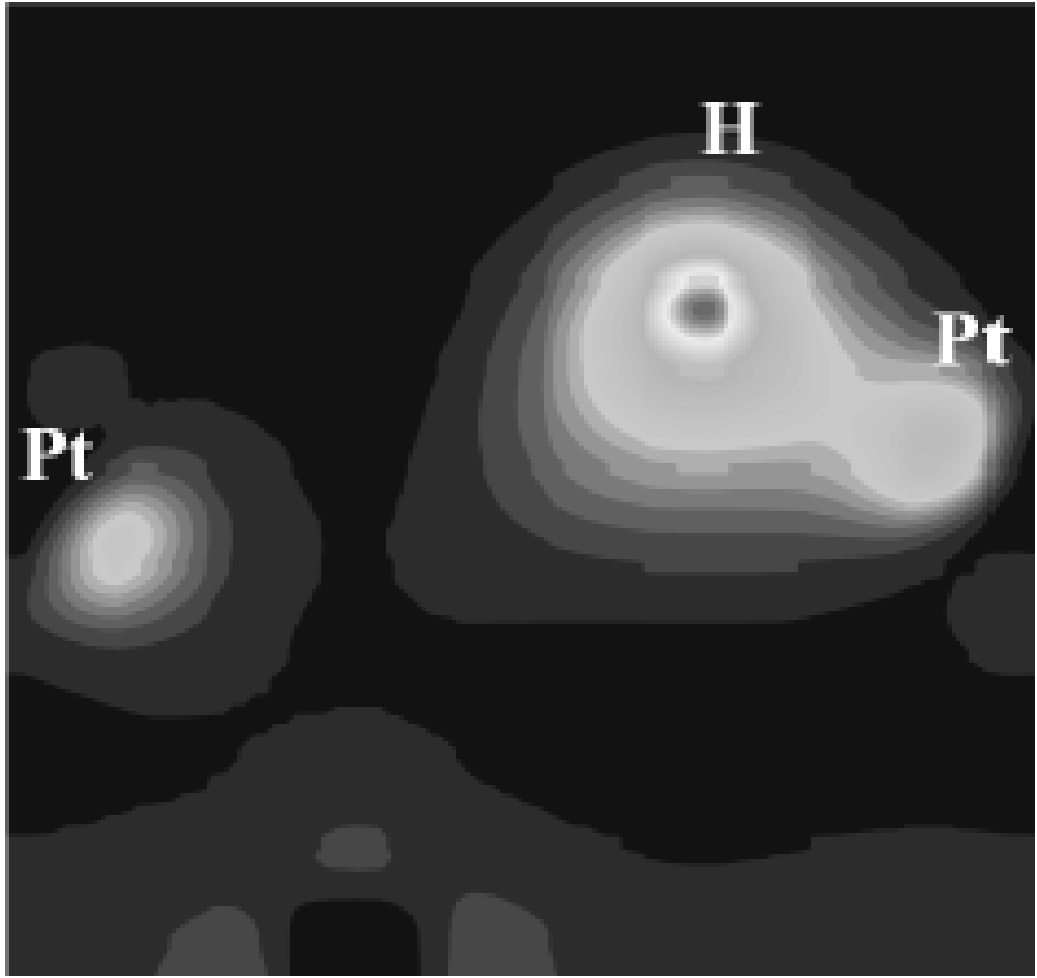

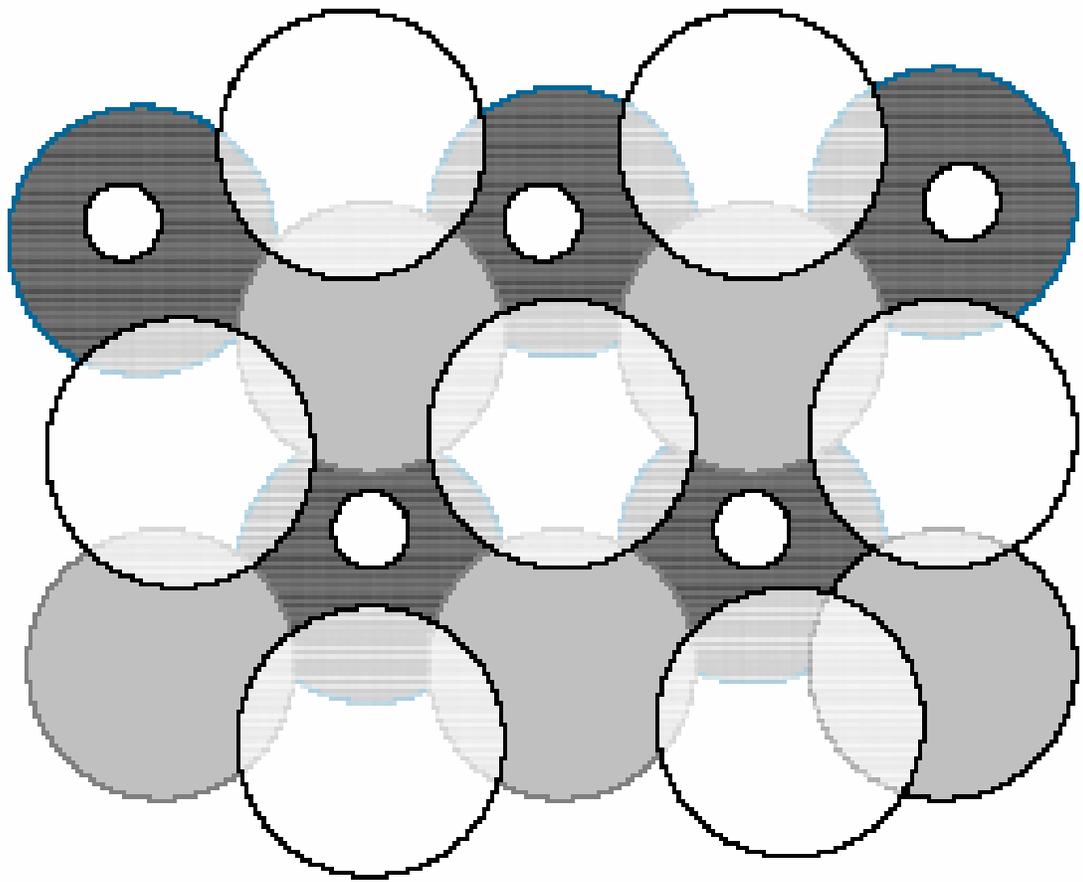

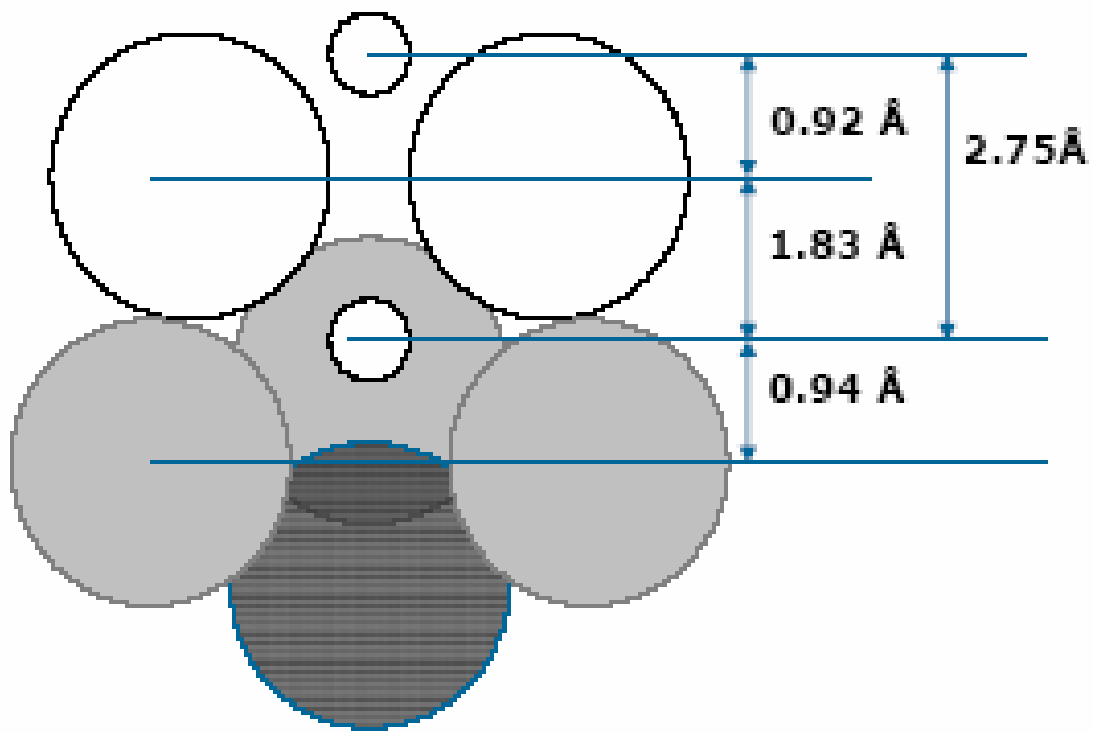